\documentstyle[epsfig]{aipproc}

\newcommand{\beq}{\begin{equation}}
\newcommand{\eeq}{\end{equation}}
\newcommand{\beqa}{\begin{eqnarray}}
\newcommand{\eeqa}{\end{eqnarray}}

\begin{document}
\title{Chiral dynamics with strange quarks}

\author{Ulf-G. Mei{\ss}ner}
\address{Universit\"at Bonn, HISKP (Th), D-53115 Bonn, Germany}

\maketitle

\begin{abstract}
In the first part of the talk, I  review what we know (or rather do not know)
about the structure of the QCD vacuum in the presence of strange quarks.
Chiral perturbation theory allows to study reactions of pions and kaons
and to further sharpen our understanding of symmetry violation in QCD.
I review recent progress on the description of pion-kaon scattering,
in particular concerning isospin violation and the extraction of threshold and
resonance parameters from Roy-Steiner equations. In the third part,
it is shown how a unitary extension of chiral perturbation theory
leads to novel insight into the structure of the $\Lambda (1405)$.
\end{abstract}

\section*{Introduction: s quark mysteries}

\noindent The strange quark plays a special role in the QCD dynamics at the
confinement scale. Here, I will discuss some open questions
surrounding chiral dynamics with strange quarks, pertinent to the 
structure of the strong interaction vacuum as well as to the structure
of light mesons and baryons. Some of these issues are: Since 
$m_s \sim \Lambda_{\rm QCD}$, is it appropriate to treat the strange quark as
light or should it be considered heavy, as in the so--called heavy kaon effective
field theory, see \cite{Roessl,Oul,FKM}~?
Why is the OZI rule so badly violated in the scalar sector with vacuum 
quantum numbers?  One example is the reaction
$J/\Psi \to \phi \pi\pi /\bar{K} K$, which is OZI suppressed to
leading  order, but even has an additional doubly OZI suppressed contribution.
The $\pi^+ \pi^-$ event distribution shows a clear peak at the energy of
980~MeV, which is due to the $f_0$ scalar meson. This lets one anticipate that 
the dynamics of the low-lying scalar mesons and the mechanism of OZI violation
are in some way related. More generally, it is of interest to learn about the
phase structure of SU($N_c$) gauge theory at large number of flavors
$N_f\,$. In QCD, we know that asymptotic freedom is lost for $N_f
\geq 17$ but from the study of the two-loop $\beta$ function  one
expects that there is a conformal window around $N_f \simeq 6$ \cite{BaZa}.
This lets one contemplate the question whether there is already
a rich phase structure even for the transition from $N_f =2$ to
$N_f =3$~? Some lattice studies seem to indicate a strong flavor
sensitivity when going from $N_f =2$ to $N_f =4$ \cite{Col,aoki}.
As discussed by many speakers at this conference, the nature of the low--lying
scalar mesons is still very much under debate (a topic I will not entertain in detail).
In the baryon sector, there are also some ``strange'' states with non-vanishing 
strangeness. More precisely, what is the  nature  of some strange baryons like 
the $\Lambda (1405)$  or the $S_{11} (1535)$, are these three quarks states or 
meson-baryon bound  states~? The latter scenario was already contemplated many years ago 
by Dalitz and collaborators \cite{Dal} and has been rejuvenated with the advent of
coupled channel calculations using chiral Lagrangians to specify the
driving interaction. In the following, I address some of these issues.

\section*{The vacuum in the presence of s quarks}

\noindent
There are many phenomenological as well as theoretical indications that the
chiral symmetry ($\chi$S) of three--flavor QCD is spontaneously broken, 
abbreviated as 
S$\chi$SB. Now the question arises what are the order parameters of the S$\chi$SB~?
Consider  the current-current correlator between vector and axial currents,
\beq
\Pi^{ab}_{\mu\nu} (q) = i \int d^4x \, {\rm e}^{i q x} \, 
\langle 0| T\{ V_\mu^a (x)  V_\nu^b (0) -  A_\mu^a (x)  A_\nu^b (0) \} |0\rangle~.
\eeq
In the three flavor chiral limit, it can be written in terms of meson
and continuum contributions and worked out explicitly,
\beq
\Pi^{ab}_{\mu\nu} (0) = -\frac{1}{4} g_{\mu\nu} \delta^{ab} F^2(3)~.
\eeq
If $\Pi^{ab}_{\mu\nu} (0) \neq 0$, then we have  S$\chi$SB.
We have  thus identified an order parameter of spontaneous chiral 
symmetry breaking, namely the pion decay constant in the
chiral limit,
\beq
\lim_{m_u,m_d,m_s \to 0} F_\pi = F(3)~.
\eeq
Its non-vanishing is a sufficient and necessary condition
for S$\chi$SB,
\beq
\Pi^{ab}_{\mu\nu} (0) \neq 0 \leftrightarrow F(3) \neq 0 
 \leftrightarrow {\rm  S}\chi{\rm SB}~.
\eeq
Naturally, there are many other possible order parameters. Often
considered is the light quark condensate,
\beq\label{qbarq}
\langle 0| \bar{q} q |0\rangle = \langle 0| \bar{u} u |0\rangle^{(3)} 
= \langle 0| \bar{d} d |0\rangle^{(3)} = \langle 0| \bar{s} s |0\rangle^{(3)}
\equiv -\Sigma(3)~,
\eeq 
because the scalar-isoscalar operator $\bar{q}q$ mixes right- and left-handed
quark fields. As will be discussed below, the quark condensate
plays a different role than the pion decay constant. Other possible color-neutral
order parameters of higher dimension are e.g. the mixed quark--gluon condensate 
$\langle 0| \bar{q}^i \sigma^{\mu\nu} G_{\mu\nu}^\alpha T^\alpha_{ij} q_j 
|0\rangle^{(3)}$ or certain four-quark condensates  $\langle 0| (\bar{q} \Gamma_1 q)
 (\bar{q} \Gamma_2 q) |0\rangle $ with $\Gamma_i$ some Dirac operator. 
It goes without saying that
the spontaneously and explicitly broken chiral symmetry can be systematically 
analyzed in terms of an effective field theory - chiral perturbation theory
(CHPT) (or some variant thereof). We now turn to the flavor dependence of these
various order parameters. For this, consider QCD on a torus,
or in an Euclidean ($t \to -ix^0$) box of size $L\times L \times L \times L$, 
understanding of course that we have to take the infinite volume 
limit at its appropriate place. Quark and gluon fields  are then subject to certain
boundary conditions, which are anti-periodic and periodic, in order.
Analyzing the spectrum of the QCD Dirac operator, one arrives e.g. at the
Banks--Casher relation. Also, the order parameters $F^2$ and $\Sigma$ are
dominated by the IR end of the Dirac spectrum \cite{DGS}, and one therefore
expects a paramagnetic effect,
\beq
\Sigma (N_f+1) < \Sigma (N_f) ~~\sim 1/L^4~, \quad
F^2 (N_f+1) < F^2  (N_f) ~\sim 1/L^2~,
\eeq
indicating a suppression of the chiral order parameters with
increasing number of flavors.
We note that the condensate is most IR sensitive. These results are
exact, the question is now how strong this flavor dependence is or how
it can be tested or extracted from some observables.

In the standard scenario of S$\chi$SB, terms quadratic in the quark masses are
small, as has been recently confirmed for  the {\em two flavor} case
from the analysis of the BNL E865 $K_{e4}$ data \cite{E865}. If these terms
are also small in the three flavor case, the so-called Gell-Mann--Oakes--Renner
ratio $X(3)$ stays close to one,
\beq 
X(3) \equiv { 2 \hat{m} \Sigma (3) \over F_\pi^2 M_\pi^2} \sim 1~,
\eeq
with $\hat m = (m_u+m_d)/2$ the average light quark mass.
There are many successes supporting this scenario, as one example I will discuss
pion--kaon scattering in the next section.  However, there is also some information
pointing towards a more complicated phase structure (suppression of $\Sigma
(3)$), as discussed next.

\parbox{7cm}{
\begin{center}
\epsfig{figure=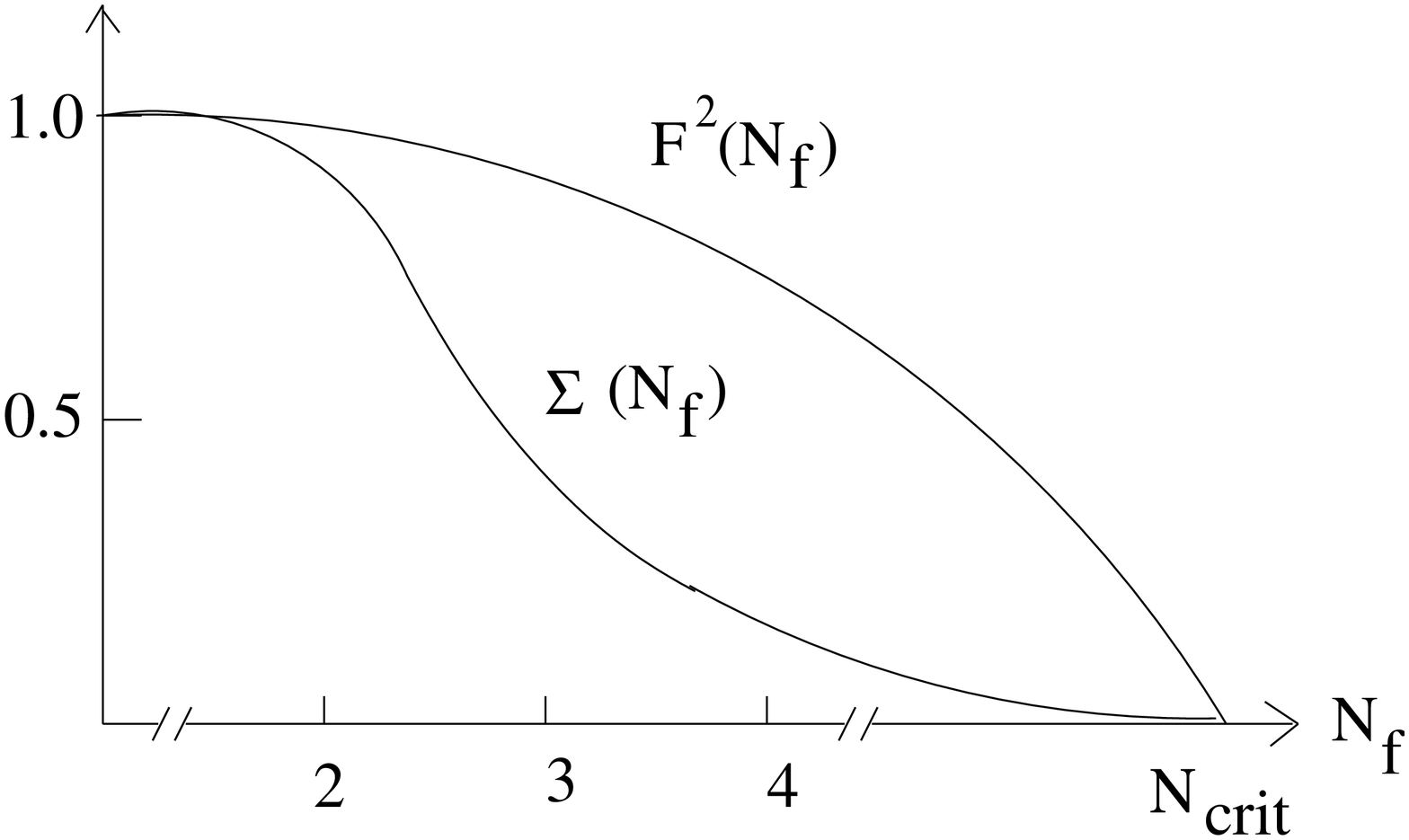,width=6.5cm,height=5cm}
\end{center}}
\parbox{7cm}{\vspace*{0.9cm}
{\small \setlength{\baselineskip}{2.6ex} {\bf Figure~1.} 
Flavor dependence of chiral symmetry breaking order parameters.
A speculative scenario with a strongly suppressed three flavor
condensate is depicted. In the standard scenario, the two lines
for the pion decay constant and the condensate would be very close
to each other.
}}
\hspace{\fill}

\noindent Moussallam \cite{BM1,BM2} investigated a sum rule for the OZI
violating correlator $\Pi_z \sim \langle \bar u u(x) \bar s s(0) \rangle_c$,
which has the form  
\beq\label{SR}
\Pi_z (m_s) = {1 \over \pi} \int_0^\infty {ds \over s} \sigma (s)~,
\eeq
and allows to relate $\Sigma (3)$ with $\Sigma (2)$.
This sum rule is super-convergent. In the approximation that the spectral function 
can be saturated by two--particle intermediate $\pi\pi$ and $K\bar K$ states, it
can be expressed entirely in terms of the (non)strange scalar form factors of the
pion and the kaon. These form factors can be calculated within CHPT, but are needed
at higher energies here (for one particular calculation in a unitarized version
of CHPT, see e.g. \cite{MO}). This gives the spectral function for energies below
$\simeq 1.6\,$GeV. One can use various  T-matrices for the $\pi\pi \to\pi\pi/K\bar K$
system 
to get an idea of the  uncertainty in this energy domain.
Above that energy, one can use pQCD. Putting all the various pieces together, one
obtains
\beq\label{Sig32}
\Sigma(3) = \Sigma (2) \, \left[ 1 - 0.54 \pm 0.27 \right]~,
\eeq
where the central value indicates a large suppression of the three
flavor condensate but the uncertainties are large enough to give
marginal consistency with the standard scenario. For a discussion of
the stability of this result  against some higher order corrections, see
\cite{BM2}. For more
investigations of such a scenario see \cite{Desc,DeSt2}.
Clearly, more work is needed to further quantify such results and to reduce
the uncertainties.

\section*{Pion-kaon scattering}
\label{sec:strange}
\noindent
Pion--kaon scattering is the simplest scattering process involving strange quarks.
Furthermore, since to one--loop accuracy all low--energy constants (LECs) are known
from other processes, one can predict e.g. the S--wave scattering lengths (given here
in the basis of total isospin 1/2 and 3/2). This has been done long time ago 
\cite{BKM1,BKM2} (in units of $M_\pi^{-1}$),
\noindent 
\begin{equation}\label{aval}
a_0^{1/2} = 0.18 \pm 0.03~[0.22\pm 0.02]~, \quad
a_0^{3/2} = -0.05 \pm 0.02~[-0.045\pm 0.008]~. 
\end{equation}
The CHPT predictions are compared to the then existing data/Roy equation analysis
in Fig.~2. Obviously,  no firm conclusion could be drawn (the dark hatched ellipse 
comes in later). 

\parbox{9cm}{
\begin{center}
\epsfig{figure=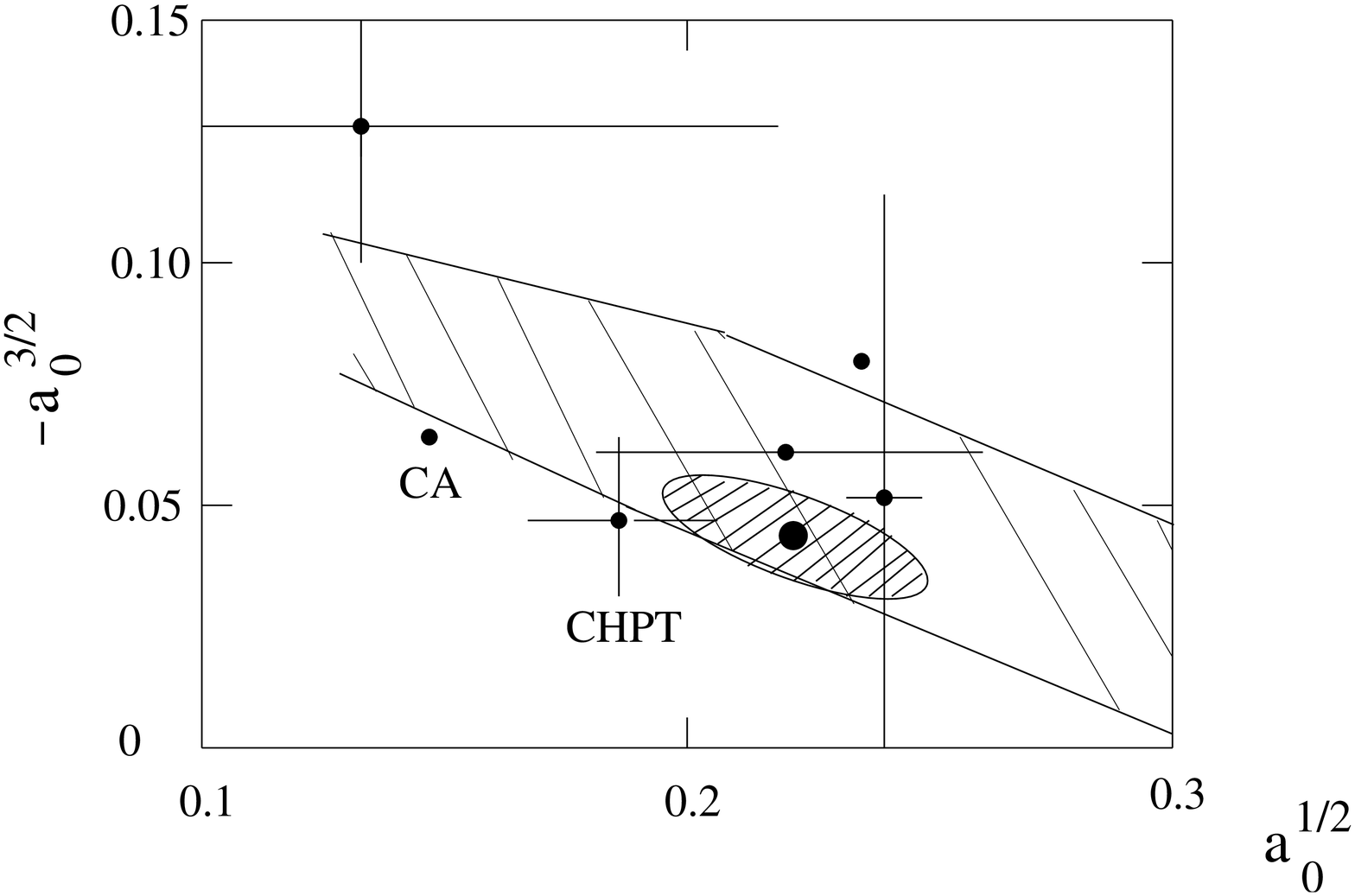,width=8.5cm,height=5.8cm}
\end{center}}
\parbox{5cm}{\vspace*{0.5cm}
{\small \setlength{\baselineskip}{2.6ex} {\bf Figure~2.} 
S-wave scattering lengths for $\pi$K scattering. The CHPT (CA) predictions
are shown by the cross (black dot). The older data/Roy equation analysis
can be traced back from \cite{BKM1,BKM2}. The dark hatched ellipse refers
to the new dispersive analysis of \cite{BDM}.
}}

\noindent
In the light of more recent and more precise data from the eighties,
that never were analyzed using dispersive methods, a novel evaluation of the 
Roy-Steiner equations was called for. This was recently achieved by
B\"uttiker et al. \cite{BDM}. They solved the Roy--Steiner
equations for the S-- and P--waves using all available input from $\pi K \to
\pi K$ and $\pi\pi \to K\bar K$ and employing Regge theory for large energies.
The outcome of this analysis are the  S-- and P--wave phase shifts below the
matching energy of 1~GeV and the amplitude in the interior of the Mandelstam
plane, in particular (sub)threshold parameters. It turns out that the
resulting phase shifts are mostly in poor agreement with the existing low energy
data, e.g. the mass of the spin--1 $K^*$ mesons from the crossing of the
P-wave isospin 1/2 phase through $\pi /2$ happens at $905 \pm 3\,$MeV, 
visibly different from the PDG value of $891.7 \pm 0.3\,$MeV. This needs 
further investigation. The resulting S--wave scattering lengths are given
in the square brackets in Eq.~(\ref{aval}), they come out consistent with the
CHPT predictions, pointing toward the validity of the standard scenario.
Similarly, the LECs extracted in \cite{BDM} agree well with earlier determinations
based on $X (3) \simeq 1$.

Another method of extracting the S-wave scattering lengths is the precise
measurements of the characteristics of pion-kaon bound states, so-called
$\pi K$ atoms. In order to relate the  lifetime and the energy shift to the
scattering lengths, one has to make use of  modified Deser formulae that include
NLO effects in isospin breaking,
\beqa
\Gamma_{\pi^0 K^0}
&\propto& \left(a_0^{3/2} -a_0^{1/2} + {\epsilon} \right)^2\, (1 +\kappa)~,
\\
\Delta E^{\rm  str}_{2S-2P}
&\propto& \left(a_0^{3/2} + 2a_0^{1/2} + {\epsilon}' \right)^2\, (1 +\kappa ')~,
\eeqa
where $\epsilon$ and $\epsilon '$ represent, respectively, the isospin 
violating corrections in the regular part of the scattering amplitudes 
$\pi^- K^+ \to \pi^0 K^0$ and  $\pi^- K^+ \to\pi^- K^+$ at threshold,
while $ \kappa \,(\kappa ')$ is an additional contribution only calculable
within the bound state formalism. There are two sources of isospin violation,
the strong contributions $\propto m_u-m_d$ and electromagnetic contributions
$\propto \alpha = e^2/4\pi$. It is most efficient to collect these two 
small parameters as $\delta \in \{m_u-m_d, \alpha\}$ and expand the corrections 
to order $\delta$ in all channels. To one--loop accuracy,
$\epsilon$ and $\epsilon '$ have been calculated in \cite{KM1,KM2} (see
also \cite{NT,N})
\beqa
a_0 (”\pi^- K^+ \to \pi^0 K^0) &=& -\sqrt{2} a_0^- \, \biggl\{ (1.\pm 0.8\%) +
\underbrace{(1.3\pm 0.1)\%}_{O(m_u-m_d)} + \underbrace{(0. \pm 1.1)\%}_{O(\alpha)}
\biggr\}~, \\
a_0 (”\pi^- K^+ \to \pi^- K^+) &=& ( a_0^- +a_0^+) \, \biggl\{ (1.\pm 16.1\%) +
\underbrace{0.2\%}_{O(m_u-m_d)} + \underbrace{(0.9 \pm 3.2)\%}_{O(\alpha)}
\biggr\}~,
\eeqa
where we have switched to the isospin basis for the scattering amplitudes,
$T^+ = (T^{1/2}+2T^{3/2})/3$ and $T^- = (T^{1/2}-T^{3/2})/3$. All quoted errors
for the different contributions are due to the uncertainties in the respective
strong and electromagnetic LECs. Note further that to leading order, the
isospin violating corrections to the elastic scattering length are entirely
given in terms of the pion mass difference, and thus are of electromagnetic origin. 
From the above equations, we can give rather precise predictions for 
 $\epsilon$ and $\epsilon '$, 
\beq
\epsilon = 1.3 \pm 1.2~\%, \quad \epsilon ' = 1.1 \pm 3.2~\%.  
\eeq
We can thus conclude that the extraction of the strong scattering amplitudes
at threshold from the lifetime and level shift of the $\pi K$ atoms is sufficiently
well under control. The isospin breaking effects in both cases are only of the
order of 1\% with an uncertainty of 1\% and 3\%, respectively. What remains
to be done is a equally precise calculation of the bound state corrections
$\kappa$ and $\kappa '$ \cite{GS}.

\section*{The nature of the $\Lambda$(1405)}

\noindent 
In this section, I will discuss some issues in the framework of SU(3) baryon
chiral perturbation theory and extensions thereof. First, it is often stated
that three flavor baryon CHPT does not converge due to the large kaon mass
and/or unitarity corrections. While that is true in certain cases, there
are many examples where indeed one can make precise predictions. As one
particular example, let me consider the charge radii of the ground state
baryon octet. To fourth order (complete one--loop calculation), the charge
radii can be can given in terms of two LECs. These parameters can be fixed from
the well measured proton and neutron electric radii, so that predictions for
the other members of the octet emerge. On the other hand, the radius of the
$\Sigma^-$ can be obtained by scattering a highly boosted hyperon beam off 
the electronic cloud of a heavy atom (elastic hadron--electron scattering). 
Such an experiment has been first carried out at CERN, demonstrating the 
feasibility of the method and later repeated with much better accuracy at FNAL.
The theoretical prediction (published before the data came out) compares
well with the result from the SELEX collaboration,
\begin{equation}
\langle r^2_{\Sigma^-}\rangle_{\rm th} 
= 0.67\pm 0.03~{\rm fm}^2~\cite{KM}, \quad
\langle r^2_{\Sigma^-}\rangle_{\exp} 
= 0.61\pm 0.12\pm 0.09~{\rm fm}^2~\cite{SELEX}.
\end{equation}
For a more detailed discussion of the status of SU(3) baryon CHPT, see
e.g. \cite{UGMioffe}.

\noindent
Next, I discuss $K^- p$ scattering. For this process, a purely perturbative treatment
is not possible (for an explicit demonstration, see \cite{NK}) 
due to the strong channel couplings and the appearance of a subthreshold
resonance, the $\Lambda (1405)$, which is supposed to be a meson-baryon bound state
rather than a genuine 3-quark state. First speculations about its possible unconventional
nature date back to \cite{Dal}. Since then many (QCD-inspired) models
have been considered, but the first work of supplementing coupled
channel dynamics with chiral Lagrangians which allows to dynamically
generate the $\Lambda (1405)$ was reported in \cite{Munich},
see also \cite{Valencia} and the review \cite{OOR}. 
A  non-perturbative resummation scheme is
mandatory to generate a bound state or a resonance. There
exist many such approaches, but it is possible and mandatory to link 
such a scheme tightly
to the chiral QCD dynamics. Such an improved approach was
developed for pion--nucleon \cite{MeOl1} and later applied to $\bar K$N
scattering~\cite{MeOl2}. 
The starting point is the T--matrix for any partial wave,
which can be represented in closed form if one neglects for the moment
the crossed channel (left-hand) cuts (for  more explicit details, see
\cite{MeOl1})
\begin{equation}
T = \left[ \tilde{T}^{-1} (W) + g(s) \right]^{-1}~,
\end{equation}
with $W = \sqrt{s}$ the cm energy (note that the analytical structure is much
simpler when using $W$ instead of $s$).
$\tilde{T}$ collects all local terms and poles (which can be
most easily  interpreted in the large $N_c$ world) and $g(s)$ is the
meson-baryon loop function (the fundamental bubble) that is resummed by e.g.
dispersion relations in a way to exactly recover the right-hand
(unitarity) cut contributions. The function $g(s)$ needs
regularization, this can be best done in terms of a subtracted
dispersion relation and using dimensional regularization.
It is important to ensure that in the low-energy
region, the so constructed T--matrix agrees with the one of CHPT (matching).
In addition, one has to recover the contributions from the left-hand
cut. This can be achieved by a hierarchy of matching conditions, 
\begin{eqnarray}
{\cal O}(p) &:& \tilde{T}_1 (W) = T_1^{\chi} (W) ~, \nonumber \\ 
{\cal O}(p^2) &:& \tilde{T}_1 (W) + \tilde{T}_2 (W) =  T_1^{\chi} (W)
+ T_2^{\chi} (W) ~, \nonumber \\
{\cal O}(p^3) &:& \tilde{T}_1 (W) + \tilde{T}_2 (W)+ \tilde{T}_3 (W) 
=  T_1^{\chi} (W)+ T_2^{\chi} (W)  \nonumber \\
&& \qquad\qquad\qquad\qquad\qquad\qquad +\, T_3^{\chi} (W) +  \tilde{T}_1
(W) \, g(s) \,  \tilde{T}_1 (W)~,
\end{eqnarray}
and so on. Here, $T_n^{\chi}$ is the T--matrix calculated within
CHPT to ${\cal O}(p^n)$. Of course, one has to avoid double counting as
soon as one includes pion loops, this is achieved by the last term in
the third equation (loops only start at third order in this case).
In addition, one can also include resonance fields by saturating the
local contact terms in the effective Lagrangian through explicit meson and
baryon resonances (for details, see \cite{MeOl1}). In particular,
in this framework one can cleanly separate genuine quark resonances
from dynamically generated resonance--like states. The former require
the inclusion of an explicit field in the underlying Lagrangian, whereas
in the latter case the fit will arrange itself so that the couplings to
such an explicit field will vanish. It was observed in ”\cite{MeOl2} that there
are indeed two poles close to the nominal $\Lambda (1405)$ resonance, as earlier
found in the cloudy bag model \cite{CB}, and later 
confirmed in \cite{Jetal,Getal}. The physics behind these two poles was recently
revealed in \cite{JOORM}. Starting from an SU(3) symmetric Lagrangian to couple
the meson octet to the baryon octet (in that limit, all
octet Goldstone boson masses and all octet baryon masses are equal), 
one could in principle generate a variety
of resonances according to the SU(3) decomposition,
\begin{equation}
 8 \otimes 8=1\oplus 8_s \oplus 8_a \oplus 10 \oplus \overline{10} \oplus 27~.
\end{equation}
As it turns out, the leading order transition potential is attractive only in the
singlet and the two octet channels, so that one a priori expects a singlet
and two octets of bound states. However, the two octets come out degenerate,
see Fig.~3. This has no particular dynamical origin but rather is a consequence of
the actual values of the SU(3) structure constants.  In the real world, there
is of course SU(3) breaking of various origins. This was parameterized in
\cite{JOORM} in terms of a symmetry breaking parameter $x$ in the
expressions for  the meson  $M_i$ and baryon masses
$m_i$ as well as the subtraction constants $a_i$ via
$M_i^2 (x) = M_0^2 + x (M_i^2-M_0^2)~,\
m_i (x) = m_0 + x (m_i -m_0)$ and $a_i (x) = a_0 + x (a_i -a_0)$, 
with $M_0 = 368\,$MeV, $m_0 = 1151\,$MeV and $a_0 = -2.148$, where
$0\leq x \leq 1$. The motion of the various poles in the complex 
energy plane as a function of $x$ is
shown in Fig.~3. We note that the two octets split, in particular, one
moves to lower energy ($I=0, 1426\,$MeV) close to the position of the singlet
($I=0, 1390\,$MeV). These are the two poles which combine to give the $\Lambda (1405)$
as it appears in various reactions.

\parbox{8cm}{
\begin{center}
\epsfig{figure=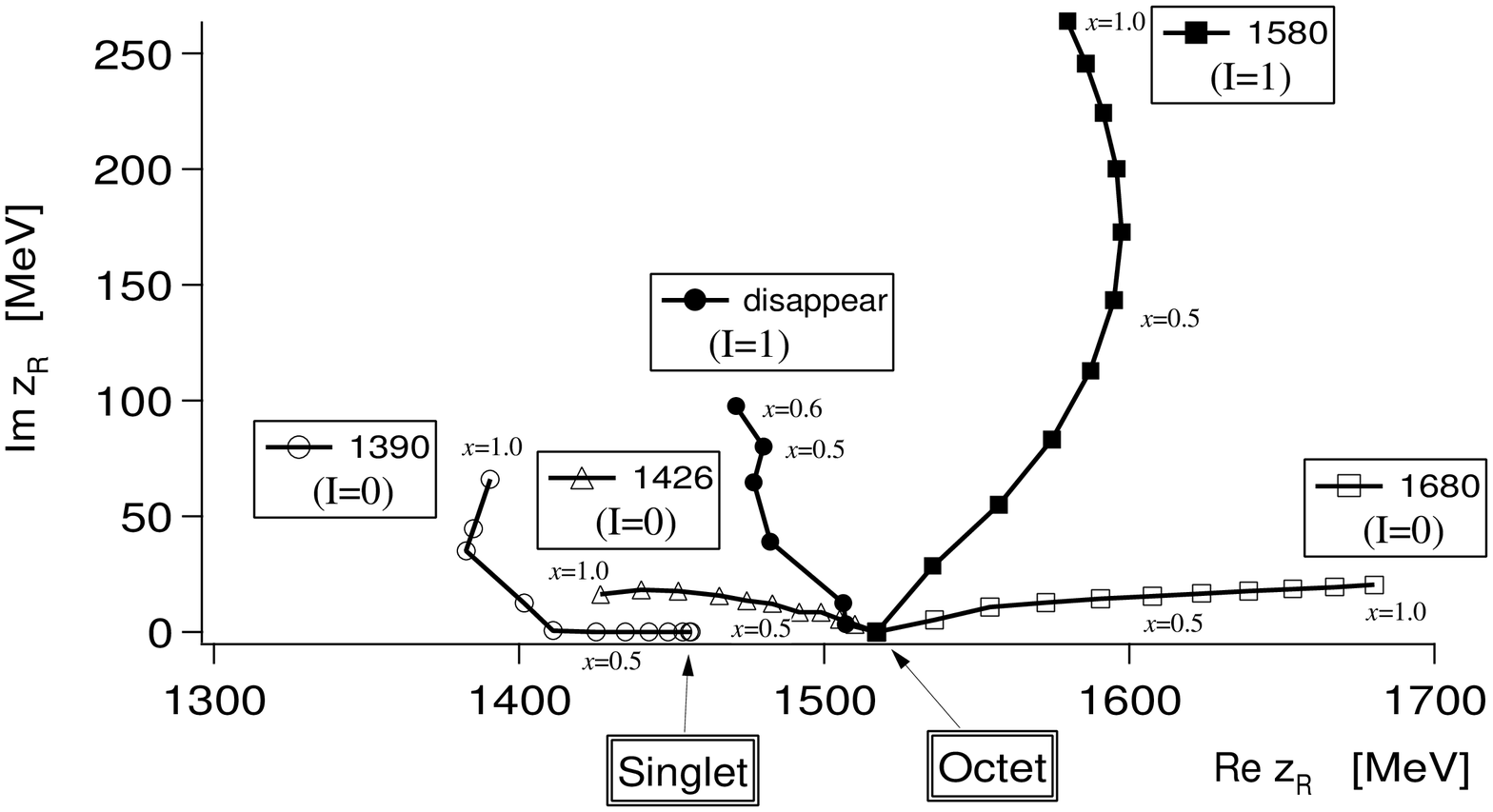,width=7.5cm,height=5.5cm}
\end{center}}
\parbox{6cm}{\vspace*{1.1cm}
{\small \setlength{\baselineskip}{2.6ex} {\bf Figure~3.} Trajectories of the poles in
 the scattering amplitudes obtained by changing the SU(3) breaking parameter $x$
 gradually. In the SU(3) limit ($x=0$), only two poles appear, one corresponding 
 to the singlet and the other to the two degenerate octets. The symbols correspond
 to the step size $\delta x = 0.1$.   
}}

\noindent
The question is now how these two different poles can actually be disentangled in
experiments? For that, one has to determine the couplings of these resonances to the
physical states by studying the amplitudes close to the pole and identifying them
with $T_{ij} = g_i g_j /(z-z_R)$ where $z_R$ is the pole position and the $g_i$ are
in general complex numbers. As shown in \cite{JOORM}, in the $I=0$ channel
the first resonance couples more strongly to $\pi \Sigma$ while the second one
has a stronger coupling to the $\bar KN$ channel. 
We thus conclude that there is not just one single $\Lambda(1405)$
 resonance, but {\em two}, and that what one sees in experiments is a
{\em superposition} of these two states.  Then, in the case  that the
 $\Lambda (1405)$ is produced from the  $\bar KN$ initial state, the peak is narrower 
as if it were produced  from an $\pi \Sigma$ initial state.
Therefore it is clear that, should there be a reaction which forces the
initial channels to be $\bar{K}N$, then this would give more
weight to the second resonance and hence produce a
distribution with a shape corresponding to an effective resonance
narrower than the nominal one and at higher energy. Such a case
indeed occurs in the reaction $K^- p \to \Lambda(1405) \gamma$
studied theoretically in Ref.~\cite{nacher}.  It was shown there
that since the $K^- p$ system has a larger energy than the
resonance, one has to lose energy emitting a photon prior to the
creation of the resonance and this is effectively done by the
Bremsstrahlung from the original $K^-$ or the proton.  Hence the
resonance is initiated from the  $K^- p$ channel
and leads to a peak structure in the invariant mass distribution 
which is narrower and appears at higher energies than the experimental
$\Lambda(1405)$  peaks observed in hadronic experiments performed so far.
Experiments of producing the $\Lambda(1405)$ with (real or virtual) photons
have been performed or are underway or will be done at SPRING-8, JLab and ELSA.
Clearly, these should be able to verify (or falsify) the two pole nature of this
particular baryon resonance. In the coupled channel approach matched to CHPT,
there is also an interesting enhancement of the $I=1$
amplitudes in the vicinity of the $\Lambda (1405)$. 
Independently of whether this  can be interpreted 
as a  resonance or as a cusp, the fact that the strength of the $I=1$
amplitude  around the
$\Lambda(1405)$ region is not negligible should have consequences for 
reactions producing $\pi \Sigma$ pairs in that region.
This has been illustrated for instance in \cite{nacher1}, 
where the photoproduction of the $\Lambda(1405)$ via the reaction 
$\gamma p \to K^+ \Lambda(1405)$ was studied. It was shown there that
the different sign in the $I=1$ component of the $\mid \pi^+ \Sigma^-\rangle$, 
$\mid \pi^- \Sigma^+\rangle$
states leads, through interference between the $I=1$ and the dominant $I=0$
amplitudes, to different
cross sections in the various charge channels, a fact that has been
confirmed experimentally very recently \cite{ahn}.

\section*{Concluding remarks}

\noindent  There are many fascinating open problems in the large field
of chiral dynamics with strange quarks. I have addressed three particular
recent issues here and refer the reader to \cite{Ulftsl} for a much broader
exposition. Certainly, one of the most important projects in the near future
is to combine the chiral coupled channel dynamics with covariant quark models such
as the Bonn one (see \cite{Bquark} and references therein)
 to solve the outstanding problem of the strong decay widths 
in such type of models and to get a better handle on the true nature of a variety
of meson and baryon resonances, which have been one of the central issues of this
conference.

\section*{Acknowledgements}
I thank  V\'eronique Bernard, Daisuke Jido, Bastian Kubis, 
Jos\'e Antonio Oller, Eulogio Oset and Angels Ramos
for enjoyable collaborations, Paul B\"uttiker and Jan Stern for useful
communications and the organizers for their hospitality.


\end{document}